# Mathematical proof of fraud in Russian elections unsound

Mikhail Simkin

In crowds, it is stupidity and not mother wit that is accumulated.

Gustave Le Bon, The Crowd: A Study of the Popular Mind

On Saturday, December 10, thousands-strong crowds swarmed Russian cities. They protested alleged fraud in December 4 parliamentary elections. Some of the allegations have mathematical origin. The Washington Post article[1] states

> "Obviously, he [Putin] doesn't agree with Gauss," one commenter wrote, referring to pioneering mathematician Carl Friedrich Gauss, who lived 200 years ago. Disenchanted Russians argue that United Russia's reported election results are so improbable as to violate Gauss's groundbreaking work on statistics.

The article does not say what exactly the problem with the election result is and what work of Gauss is relevant. It only says that he lived 200 years ago. However, this may be enough to trigger an alert: the science had somehow advanced during past 200 years. I took a closer look at the allegations.

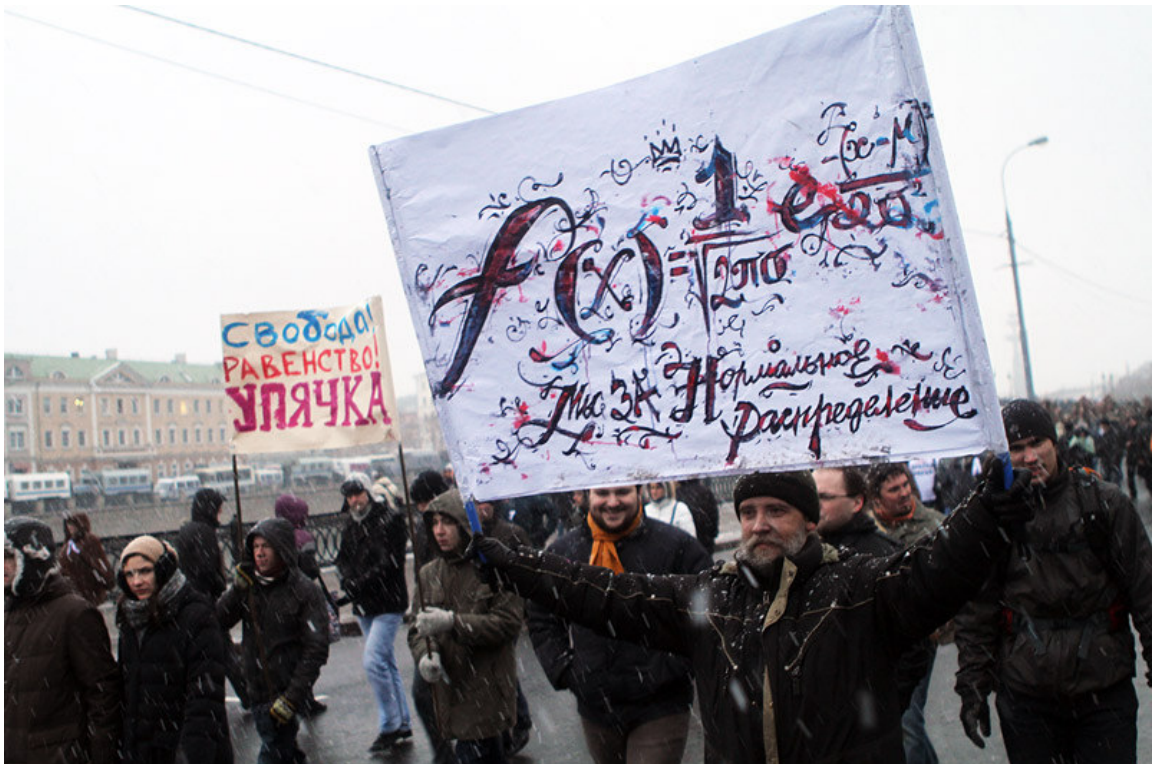

The banner says "We are for Normal distribution."
(Image from http://nl.livejournal.com/1082778.html )

---

[1] http://www.washingtonpost.com/world/europe/russians-scoff-at-medvedev-election-inquiry/2011/12/11/gIQAmBR8nO_story.html

In the article[2] titled "Mathematics against Election Committee: Gauss against Churov [the head of the committee]" the blogger complained that the distribution of the percent of the vote for the United Russia party among election precincts is non-Gaussian. This, he wrote, is an evidence of an election fraud because Gaussian distribution arises

> Always. In every case, when there is not one factor, but many. Whatever is measured in large quantities. Make a plot of how many millions of men in the country have the height of 165, 170, 175 centimeters and so on – and you will also get a symmetric bell-curve with the top corresponding to the most typical height in the country.

Yes, the heights of people are Gaussian-distributed. But what about incomes? They are distributed as if indeed most people were 170 centimeters tall, but often you would meet a three-meter guy. Rarely you would encounter a five-meter man, more rarely – a ten-meter one. Sometime, from a distance, you would see a hundred-meter person. And there would be several hundred-kilometer chaps in the country. This distribution is very far from Gaussian, but for some reason it does not attract the wreath of our mathematicians, of our Berezovskies[3]. There are many non-Gaussian distributions in both nature and society [1] and there is no reason to believe that the distribution of the percent of the vote for a party among election precincts must be Gaussian. We can illustrate this using the work of the Russian mathematician Andrei Markov [2].

The culprit is that to get a Gaussian distribution the many factors must be independent. Consider an urn with one white and one black ball. Let us pull a random ball out of the urn, record its color and put the ball back. If we make a large number of such independent trials, the distribution of the fraction of the pulled white balls will be Gaussian. This is because the color of the ball we pull this time does not depend on the color of the ball we pulled out in the preceding trial: that very independence of factors. In the case of elections independence of factors means, that people chose their political views independently of their neighbours, co-workers and friends. To account for dependent events Markov [2] modified the model in the following way. The urn initially contains one white and one black ball. We pull out a random ball, then put it back and in addition add to the urn another ball of the same color. After two trials, we could pull out either two black, two white or one black and one white ball. Elementary combinatorics shows that these three combinations are equally probable. That is the number of pulled out white balls can be 0, 1, or 2, and each of these numbers has the same probability – 1/3. You can prove by induction that after $N$ trials all numbers from 0 to $N$ of pulled out white balls are equiprobable (you can also find the proof in Chapter 7 of Ref. [1]). Interestingly, this uniform distribution even somewhat resembles the mathematically impossible distribution presented on the banner of protest (next page).

---

[2] http://www.newsland.ru/news/detail/id/838730/
[3] http://www.significancemagazine.org/details/webexclusive/1393253/Berezovsky-number.html

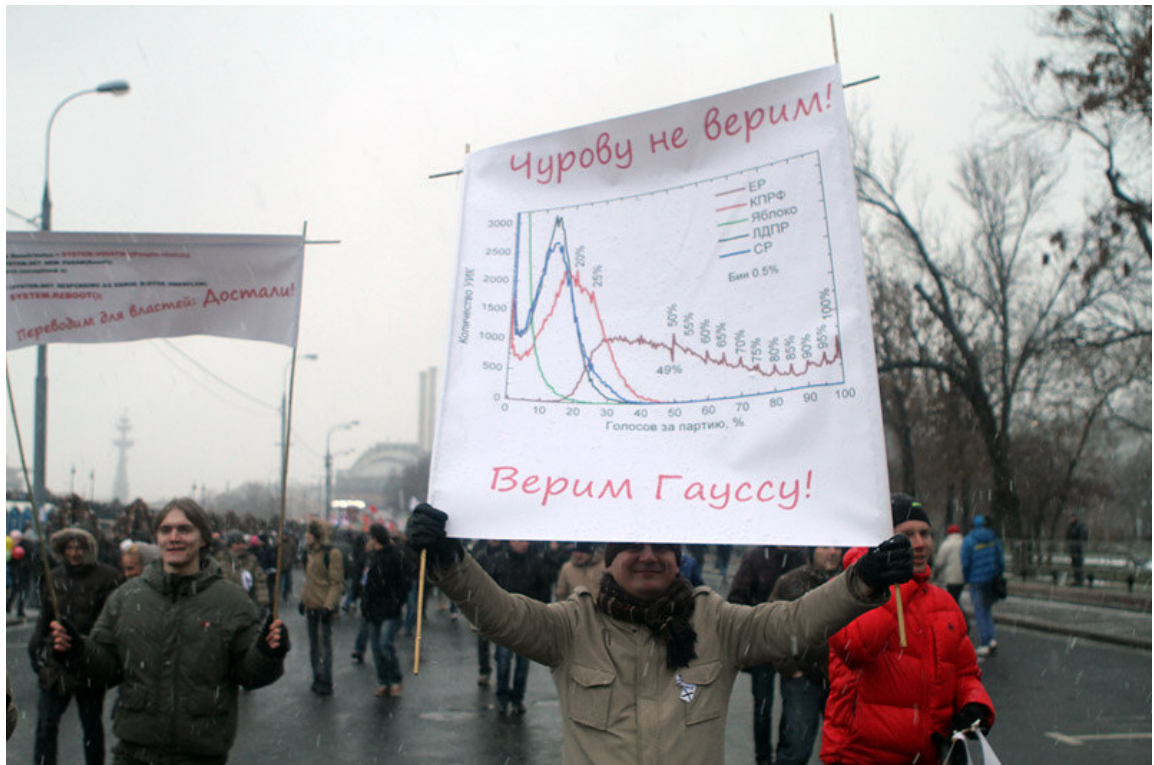

The banner says "We do not trust Churov [the head of Election committee]! We trust Gauss" (Image from http://nl.livejournal.com/1082778.html )

What relation can the ball problem have to the elections? Let us consider the following model. In a small city, which has only one election precinct, in the beginning there are two party members. One represents the White ball party and the other – Black ball party. Each of them starts agitating for his party. When the agitator persuades someone to join his party, the new party member himself starts agitating. Let us suppose that the agitator for White ball party got lucky the first. Now there are two people agitating for White party and only one for Black party. If we suppose that each agitator has equal chances to succeed, then the probability that the new party member will join the White party is two times bigger than the probability that he joins the Black party. We have a one to one correspondence with Markov's model. This means that the vote percentage distribution among the precincts must be not Gaussian, but uniform. Of course, the model we just considered is oversimplified. We completely neglected the influence of people living in different precincts on each other. This dependence, though lesser than the influence of neighbours, co-workers and friends still exists. That we have only two parties in the model may be not that big of a defect. The bloggers agitated to vote for anyone, but the United Russia party. So we can present the situation like a choice of either for or against United Russia. Of course, there is no reason to believe that the model I just described is close to reality. It is thus not clear what the distribution of vote percentages among precincts must be according to Science. It is, however, clear that there is no grounds for a demand for this distribution to be Gaussian.